\documentclass[graybox]{svmult}

\usepackage{type1cm}        
\usepackage{makeidx}         
\usepackage{graphicx}        
\usepackage{multicol}        
\usepackage[bottom]{footmisc}

\usepackage{newtxtext}       %
\usepackage{newtxmath}       

\makeindex             

\usepackage{float}
\usepackage{algorithm}
\usepackage{algorithmic}
\usepackage{multirow}
\usepackage{hhline}
\usepackage[caption=false,font=footnotesize,labelfont=sf,textfont=sf]{subfig}
\usepackage[breaklinks, colorlinks,citecolor=blue, bookmarks=false]{hyperref}
\hypersetup{ bookmarksnumbered=true,colorlinks=true, citecolor=red, bookmarksopen=true}
\usepackage[activate={true,nocompatibility},final,tracking=true,kerning=true,spacing=true,factor=1100,stretch=10,shrink=10]{microtype}

\begin{document}

\title*{ExerSense: Real-Tme Physical Exercise Segmentation, Classification, and Counting Algorithm Using an IMU Sensor}
\titlerunning{Real-Tme Physical Exercise Segmentation, Classification, and Counting Algorithm}
\author{Shun Ishii, Kizito Nkurikiyeyezu, Anna Yokokubo and Guillaume Lopez}

\institute{Shun Ishii \at Aoyama Gakuin University, Sagamihara, Japan \email{sishii@wil-aoyama.jp}}
%
%
\authorrunning{Ishii et al. (2020)}
\maketitle
\vspace{-2cm}
\abstract{Even though it is well known that physical exercises have numerous emotional and physical health benefits, maintaining a regular exercise routine is quite challenging. Fortunately, there exist technologies that promote physical activity. Nonetheless, almost all of these technologies only target a narrow set of physical activities (e.g., either running or walking but not both) and are only applicable either in indoor or in outdoor environments, but do not work well in both environments. This paper introduces a real-time segmentation and classification algorithm that recognizes physical exercises and that works well in both indoor and outdoor environments. The proposed algorithm achieves a 95\% classification accuracy for five indoor and outdoor exercises, including segmentation error. This accuracy is similar or better than previous works that handled only indoor workouts and those use a vision-based approach. Moreover, while comparable machine learning-based approaches need a lot of training data, the proposed correlation-based method needs one sample of motion data of each target exercises.}
\section{Introduction}
Exercise and physical activity have well-documented mental and physical health benefits \cite{Dahn2005, Warburton2006}. People with regular physical activity are healthier and have a better mood. They are also less prone to several chronic diseases (e.g., cardiovascular disease, diabetes, cancer, hypertension, obesity, and depression) and live much longer compared to those with a sedentary lifestyle. Consequently, active daily living is recommended to all people of all ages \cite{Warburton2006}. Unfortunately, despite the numerous benefits of regular physical activity, it is challenging for most people to stay motivated and keep adherence to a regular workout schedule \cite{Dishman1985}. Indeed, people easily lose self-motivation. Additionally, at least for beginners, proper physical exercise necessitates training.
\par
Researchers and exercise therapists have proposed numerous strategies that help improve adherence to a regular exercise schedule \cite{Tuso2015}. These include, among other things, encouraging people to be physically active and to create an environment that makes it easier for people to be physically active in their homes. For example, \cite{Dharia2018} use smartphone data and developed a fitness assistant framework that automatically generates a fitness schedule. The framework also incorporates social interaction to increase the engagement of its users. The most advanced state of the art aims at serving as a substitute for a personal trainer. For instance, FitCoach \cite{Guo2020} is a virtual fitness coach that uses wearable devices and assesses the patterns and position of its users during workouts in order to help them achieve an effective workout and to prevent them from workout injuries. Extensive experiments, both in indoor and outdoor conditions, have shown that FitCoach can assess its users' workout and provide an adequate recommendation with accuracy $> 90\%$.
\par
This paper introduces a method that utilities an inertial measurement unit (IMU) sensor in order to provides an accurate real-time segmentation, classification, and counting of physical exercises. The proposed method works well both indoor and outdoor environments and it only uses one wearable sensor. Recently, IMU sensors have become more widely adopted for physical activity recognition \cite{Prakash2019, Wahjudi2019, Hausberger2016}. Some IMU-based systems (e.g., \cite{Prakash2019}) are used for step counting and walking detection to encourage its users to increase their ambulatory physical activity. Other methods (e.g., \cite{Wahjudi2019}) automatically recognize various walking workouts (e.g., walking and brisk-walking). Finally, advanced IMU-based systems (e.g.,\cite{Oreilly2018, Crema2017}) aim at altogether bypassing the need for personal physical trainers. They monitor their users during exercise and classify their exercises technique and provide feedback to improve their workout.
\par
The proposed approach provides practical enhancements to the existing body of research. First, most existing approaches have practical limitations. For example, methods for outdoor physical activity recognition are usually based on frequency analysis, and since the number of cycles is large, a few mis-classifications are tolerable, but such errors are not tolerable for plyometric exercise. The proposed method works well for short-term cyclic movement exercises (e.g., push-ups) and for long-term cyclic quick movements exercises (e.g., running and walking). Second, unlike other comparable machine learning-based approaches that need
a lot of training data, the proposed method needs one sample of motion data of each target exercises and yet performs reasonably well (accuracy $> 95\%$). Finally, although not yet validated, the proposed approach has also the potential to evaluate the quality of the workout.
\section{Related Work}
\subsection{Vision-based exercise recognition}
There exist many studies that quantitatively evaluate the performance of sports and physical exercises. These research are often based on three-dimensional (3-D) image analysis, whether it is for baseball \cite{feltner1986,dillman1993,werner1993,wang1995,barrentine1998,matsuo2000} or tennis \cite{emmen1985,chow2003,chiang2007,razali2012}. Typically, the evaluation is based on kinematics and the dynamics of joint motions of shoulder, elbow, forearm, wrist, and fingers during pitching. For example, Antón et al.\cite{D2015} introduced a Kinect-based algorithm for the monitoring of physical rehabilitation exercises. The algorithm recognizes the main components of the exercises, postures, and movements in order to assess their quality of execution. Moreover, this game-like immersive framework motivates them to do the rehabilitation sessions more enjoyable. Despite only a few samples in the training step, the algorithm is capable of making real-time recognition of the exercises and achieved a monitoring accuracy of 95.16\% in a real scenario when evaluated on 15 users.
\par
In general, vision-based approaches are more accurate than wearable sensor-based approaches for exercise recognition. Although they achieve good performances, the use of a vision-based sports/exercise recognition system is limited to dedicated locations. Besides, 3-D image analysis is complex and computationally intensive. This limitation, is, however, minimized by the possibility to perform some preprocessing on the sensor level.
\subsection{Recognition of movement repetition based exercises}
One of the relevant previous work is that of Dan et al. \cite{Dan2014}, who introduced RecoFit, a system for automatically tracking repetitive exercises such as weight training and calisthenics via an arm-worn inertial sensor. They addressed three challenges, segmenting, recognizing, and counting of several repetitive exercises. They achieved precision and recall greater than 95\% in segmenting exercise periods, 99\%, 98\%, and 96\% of recognition of 4, 7, and 13 exercises respectively, and 93\% of $\pm1$ repetition of counting accuracy. However, the method of RecoFit needs 5 seconds to segment and recognize exercise. In the case of a small number of counts, it cannot find correct exercise and count. Besides, it requires a dedicated device attached to the forearm, which implies a supplementary cost for users that has to buy a device for a particular and limited usage, as well as the burden of attaching a device to an unusual part of the body.
\subsection{Recognition of movements using smartwatches}
With the popularity of smartwatches and other smart wearable devices that integrate multiple sensors, there is less need for exercise-specific hardware development. Smartwatches generally have built-in microelectromechanical systems (MEMS), IMU, and pulse rate (PR) sensors. Therefore, these devices need only software applications to be developed for each targeted sport or exercise. Examples are the applications developed by Lopez et al. \cite{lopez2019} for supporting an athlete or a beginner with baseball pitching action and tennis serve action. The personal sport skill improvement support application is running on Sony's SmartWatch SWR50 and does not even need to communicate with the paired Smartphone to perform on-site movement analysis and feedback. The comparative research using the proposed Smartwatch applications for sport skill improvement support achieved very encouraging results.
\par
Viana et al. \cite{viana2018} proposed an application called GymApp, similar to the system mentioned above, but applied to workout exercise recognition. It also runs on Android OS smartwatches and monitors physical activities, for example, in fitness. It has two modes of operation: training mode and practice mode. In training mode, an athlete is advised to perform an exercise (e.g., biceps curl) with lighter weight and with the supervision of a fitness instructor, to guarantee the correctness of the performed exercise. The application then gathers sensory data and builds a model for the performed exercise, using supervised machine learning techniques. Then, in the practice mode, the recorded sensory data are compared with the previously acquired data. The application calculates the similarity distance and from the result, estimates how many repetitions of the exercise were performed correctly.
\par
Although the above-described studies are very promising, they are based on machine-learning techniques. It implies a necessary preliminary step to collect data to train a model for each type of targeted movement. This training step is a burden for the users and a disadvantage towards deploying the technology.
\subsection{Recognition of movements using an earable device}
While many researchers and developers have been developing applications based on smartphones and smartwatches, Kawsar et al. proposed and developed a new wearable platform called ``eSense" (see Figure \ref{eSense}). It consists of a pair of wireless earbuds augmented with kinetic, audio and proximity sensing. The left earbud has 6-axis IMU and a Bluetooth Low Energy (BLE) interface to stream sensor data to a paired device. Also, both earbuds are equipped with microphones.
\par
The use of earphones to listen to music while exercising is widespread, and though the eSense platform is still recent, it already attracted the attention of many research teams. It can simultaneously monitor exercises analyzing the sensory information and provide feedback that does not bother the visual field of the user through the acoustic interface. Indeed, repeated check of some visual feedback provided on a smartphone or smartwatch screen may be dangerous and the cause of accidents when done during exercises implying motion. For example, Radhakrishnan et al \cite{radhakrishnan2019}. proposed to use the eSense platform to improve user engagement during indoor weight-based gym exercises.
\subsection{Research works about step counting}
While step-count has been extensively studied in the ubiquitous computing community, false positives are still unsolved issues. The main reason for that is due to motion noise that produces the same signals as walking. However, if IMUs are at the ear, these noisy motions do not propagate up to the ear. Prakash et al. introduced the advantages of eSense in counting the number of steps of walking \cite{Prakash2019}. While head movements can still pollute this bouncing signal, they developed methods to alleviate the problem. The proposed system, STEAR, achieved 95\% step-count accuracy under the condition that smartphone and Fitbit-like systems falter. This research has excellent possibilities for wearable technologies, but it only focused on typical outdoor exercises like walking.
\par
Most of the works related to detailed exercise recognition achieve around 95\% for each defined exercise under the condition of only indoor workouts or only outdoor exercises like walking and running. So, this research aims to recognize both indoor and outdoor exercises while keeping with the same accuracy.
\par
Besides, many of them are based on machine-learning techniques, which often require a new dataset for each new user. So, this research also aims at proposing a method that provides a very accurate real-time segmentation, classification, and counting of physical exercises, without needing re-calibration for each user.
\section{Method}
\subsection{Outline of the proposed method}
\begin{figure}
     \centering
    \includegraphics[width=1\linewidth]{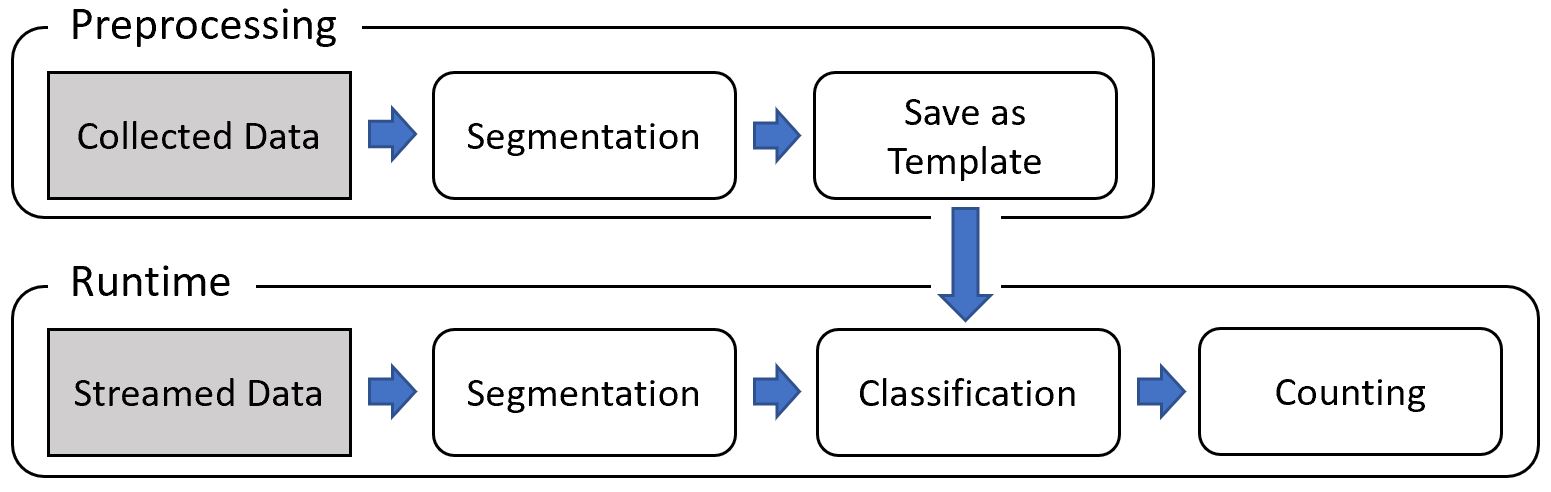}
    \caption{Schematic of the architecture of the proposed method}
    \label{method}
\end{figure}

Figure \ref{method} represents a schematic of the architecture of the proposed method. It consists of  two phases: the pre-processing and runtime phase.
\par
In the pre-processing phase, some data of the targeted exercise are collected. Only one single motion of the exercise is necessary and saved as the template 3-D acceleration signal for this exercise. That is a significant advantage of the correlation-based approach against approaches based on machine learning. Notably, in the case of exercise recognition, it is tough to collect training data for machine learning.
\par
The runtime phase starts with the segmentation of the streamed acceleration signal into single motions by finding the peaks in the synthetic acceleration signal. The next section explains in detail the segmentation process. Then, every segmented 3-D acceleration signal is classified by comparison with each exercise template produced in the pre-processing phase using a correlation-based algorithm, and the count of classified exercise is incremented.
\subsection{Segmentation algorithm for single motion extraction}
\begin{figure}
    \centering
    \includegraphics[width=0.85\linewidth]{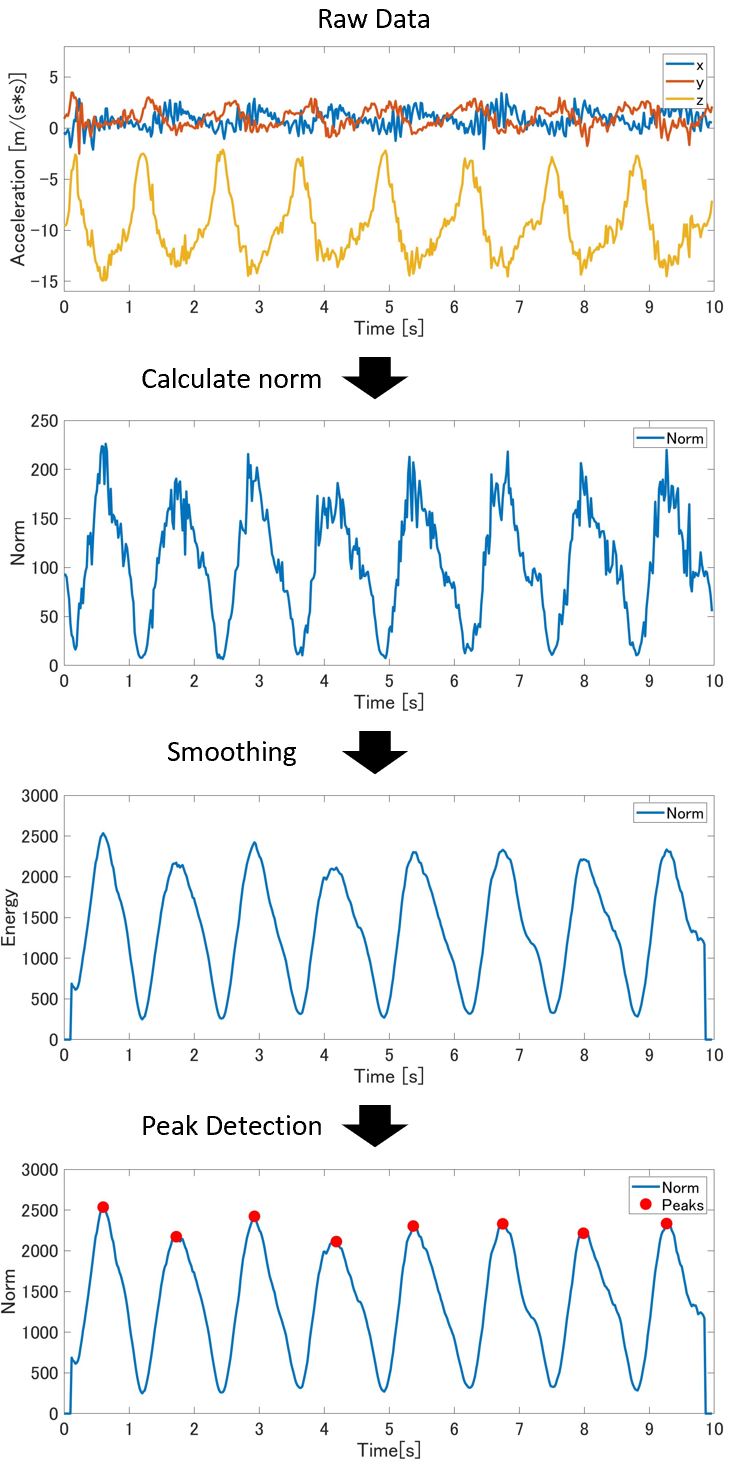}
    \caption{Example of the processing flow of the segmentation algorithm applied to 3-D acceleration signal collected during push-ups exercise}
    \label{segmentation}
\end{figure}
\par
Figure \ref{segmentation} illustrates the process of segmentation algorithm from a 3-D acceleration signal collected during push-ups exercise. First, the synthetic acceleration of streamed inertial sensor data, which is the norm of the 3-D acceleration signal, is calculated. In the case of push-ups, as shown in Figure \ref{segmentation}, peaks detection and motion segmentation may be performed using only the longitudinal acceleration (Z-axis in Figure \ref{segmentation}) of raw data. However, it is not the right solution since this research targets not only push-ups but also other types of exercise, including those that do not imply movements in the longitudinal direction. Therefore, the synthetic acceleration is more appropriate, though it presents  a disadvantage of reducing the differences between movements that are similar but along a different axis.
\par
As shown in the second plot of Figure \ref{segmentation}, the result of the norm includes much noise. Applying short-term energy, it enables not only to emphasize significant signal variations but also to smooth it as shown in the third plot. Smoothing is important to detect only motion start and end peaks easily.
\par
Then, a sliding window of 0.25 seconds length is used to detect peaks. If the center value of the window is the maximum value of the window, then it is determined as a peak. The fourth plot shows detected peaks plotted on the smoothed norm of acceleration signal collected during push-ups exercise.
\par
Finally, the 3-D acceleration signal is segmented by extracting the data between the period of two consecutive peaks.
\par
\begin{figure}
    \centering
    \includegraphics[width=1\linewidth]{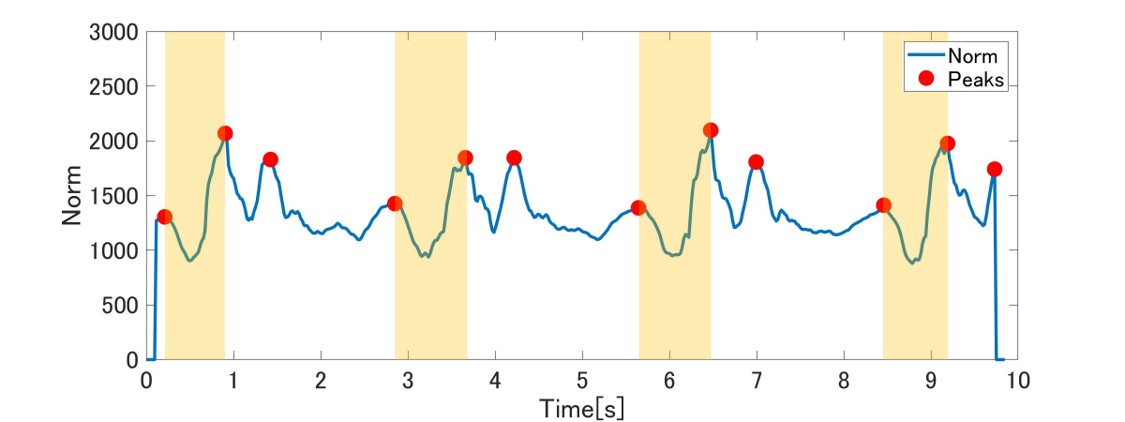}
    \caption{The smoothed synthetic acceleration signal and detected peaks during sit-ups exercise}
    \label{situpPeaks}
\end{figure}

In most cases, one peak is detected for each motion. However, in the case of sit-ups, three peaks are detected for each motion (see Figure \ref{situpPeaks}). To be able to deal with this case, one of the three peak-to-peak periods (yellow-colored in Figure \ref{situpPeaks}) is defined as sit-up base motion, and when it is detected, the latter two segments are ignored. If the person starts another exercise after sit-ups, there is no problem for counting because for motions such standing up, sitting down, and other posture changes, the smoothed synthetic acceleration also includes more than two peaks.
\subsection{Preparation of exercise templates}
In order to use the correlation-based approach, templates data of all targeted exercises are necessary. The segmentation algorithm above described was used to get one motion data of each exercise. For each exercise, one of the detected segments of motion data from a peak to the next peak was selected randomly as the template. Compared with the machine learning approach, what is needed is just doing the exercise for once or a few times to collect template data.
\subsection{Classification of extracted motion segments}
\begin{figure}
    \centering
    \includegraphics[width=1\linewidth]{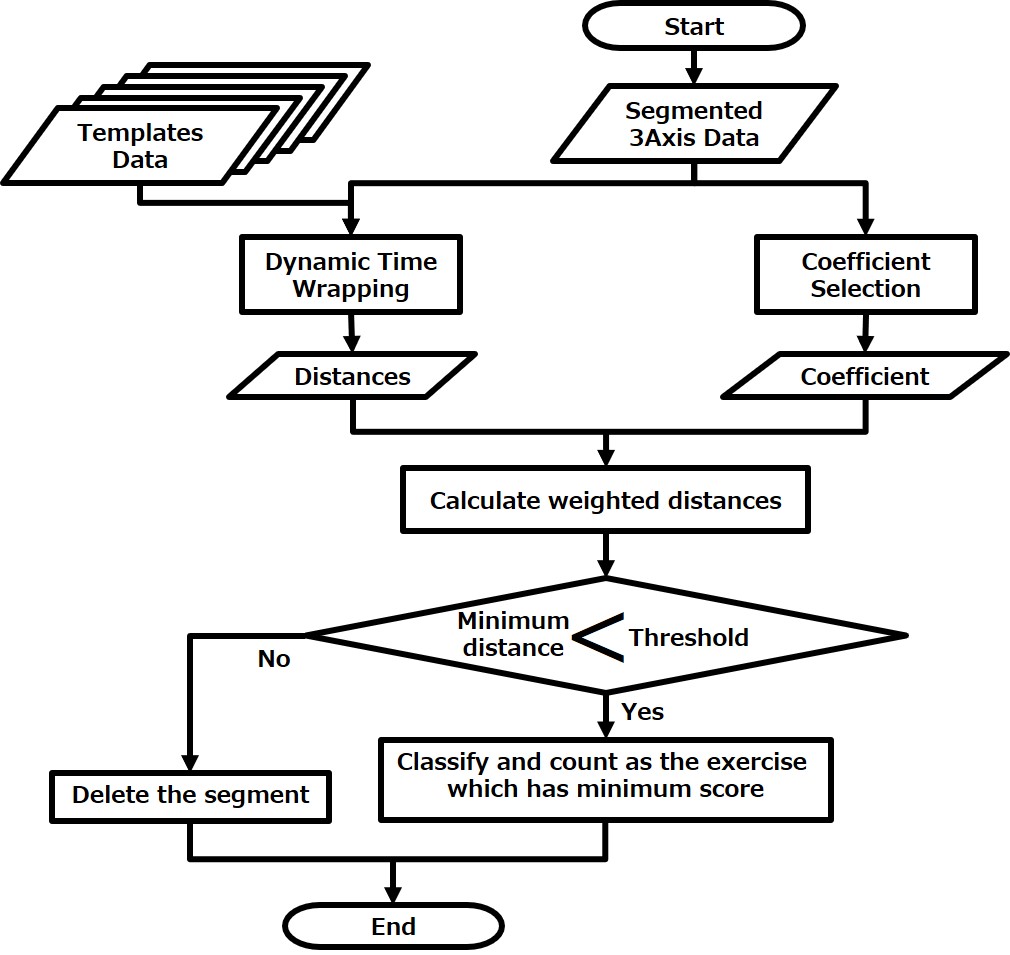}
    \caption{Flow chart of the processing flow of the proposed method to classification extracted motion signal segments}
    \label{classification}
\end{figure}
Figure \ref{classification} shows the processing flow of the proposed classification method. After extracting the 3-D acceleration signal corresponding to a single motion through the segmentation process, the raw data are used to calculate the correlation with the templates of every exercise. The Dynamic Time Warping (Algorithm \ref{DTW}) algorithm is applied to calculate the distance between every template signal and the extracted signals. The Dynamic Time Warping (DTW) can calculate the distance between two time-series data that have different lengths. This is a crucial property since it offers the capability to deal with the shape of signals issued from one identical exercise, independently of the speed the exercise motion is performed. Then, multiple weights are applied to the calculated DTW score for each exercise template. The variance of the three axes determines the weights. For example, if the maximum variance of the axis is Z-axis, the weight of ``Exercise 2" is ``0.9." The direction of maximum movement and body influence the final DTW scores, which are used for classification. It is an essential point of the proposed method. Finally, the proposed method classifies the exercise that has the minimum weighted DTW score as the ongoing exercise. However, if the minimum score is higher than a pre-defined threshold, the segment is ignored as non-exercise.
\begin{algorithm}
    \centering
    \caption{Dynamic Time Warping}
    \label{DTW}
    \begin{algorithmic}
        \REQUIRE{$n > 0 \land m > 0$}
        \STATE{$DTW \Leftarrow array[0...n, 0...m]$}
        \FOR{$i = 0 \, \ldots \, n$}
            \STATE{$DTW[i, 0] \Leftarrow infinity$}
        \ENDFOR
        \FOR{$j = 0 \, \ldots \, m$}
            \STATE{$DTW[0, j] \Leftarrow infinity$}
        \ENDFOR
        \STATE{$DTW[0, 0] \Leftarrow 0$}

        \FOR{$i = 0 \, \ldots \, n$}
            \FOR{$j = 0 \, \ldots \, m$}
                \STATE{$cost \Leftarrow ||(s[i] - t[j])||$}
                \STATE{$DTW[i, j] \Leftarrow cost + min(DTW[i-1, j], DTW[i, j-1], DTW[i-1, j-1])$}
            \ENDFOR
        \ENDFOR
        \RETURN{$DTW[n, m]$}
    \end{algorithmic}
\end{algorithm}
\subsection{Counting}
After the classification step, it is easy to count each exercise. Only what we need to do is to iterate by one the counter for each classified exercise. However, in the case of sit-ups, the proposed method divides one motion into three segments. One of the three three segments will be similar to template data but others are unlikely. So, we can count correctly with the combinations of segmentation and classification.
\section{Validation of the proposed method}
In order to validate the proposed method, experiments have been conducted for five exercises.
\subsection{Experimental conditions}
The experimental conditions to validate the proposed method are described following.
\begin{itemize}
    \item \textbf{Experimental circuit}\textemdash As mentioned in the introduction, this research targets regular exercises including indoor and outdoor. So a circuit to do five regular exercises has been created to evaluate the proposed method. The order of the five exercises, which explain in the next section, is determined randomly and systematically.
    \item \textbf{Participants} \textemdash Fifteen participants have been recruited from University students. Participants varied in weight from 58kg to 80kg, and also they self assessed  as performing exercise ``at least once a week," with an average of four times a week. Each participant performed once all exercises according to the conditions described above. Due to the missing value of six participants, we used valid data of nine participants to validate the proposed method.
    \item \textbf{Sensors}\textemdash  Though this research aims at being deployable with various types of commercially available general use wearable devices such as smartwatches, smart earphones, smart glasses, and chest bands, the study reported in this thesis is based on the signal of an IMU that is mounted to the chest to develop basic classification algorithm. One can assume that the chest movement, as the head, has less noise than other body parts. Chest sensors are also commonly used by people practicing exercise several times a week to monitor their heart rate. Nevertheless, in these experiments, three other wearable devices were also attached in the prevision of future works: a smartwatch attached to the left wrist, a smartphone attached to the upper left arm, and the eSense earbud attached to the left ear (see Figure \ref{sensors}). Future works will be to develop an accurate and robust exercise classification algorithm and application software that can be deployed for various wearable devices that most people have nowadays.
    \begin{figure}
        \includegraphics[width=1.0\linewidth]{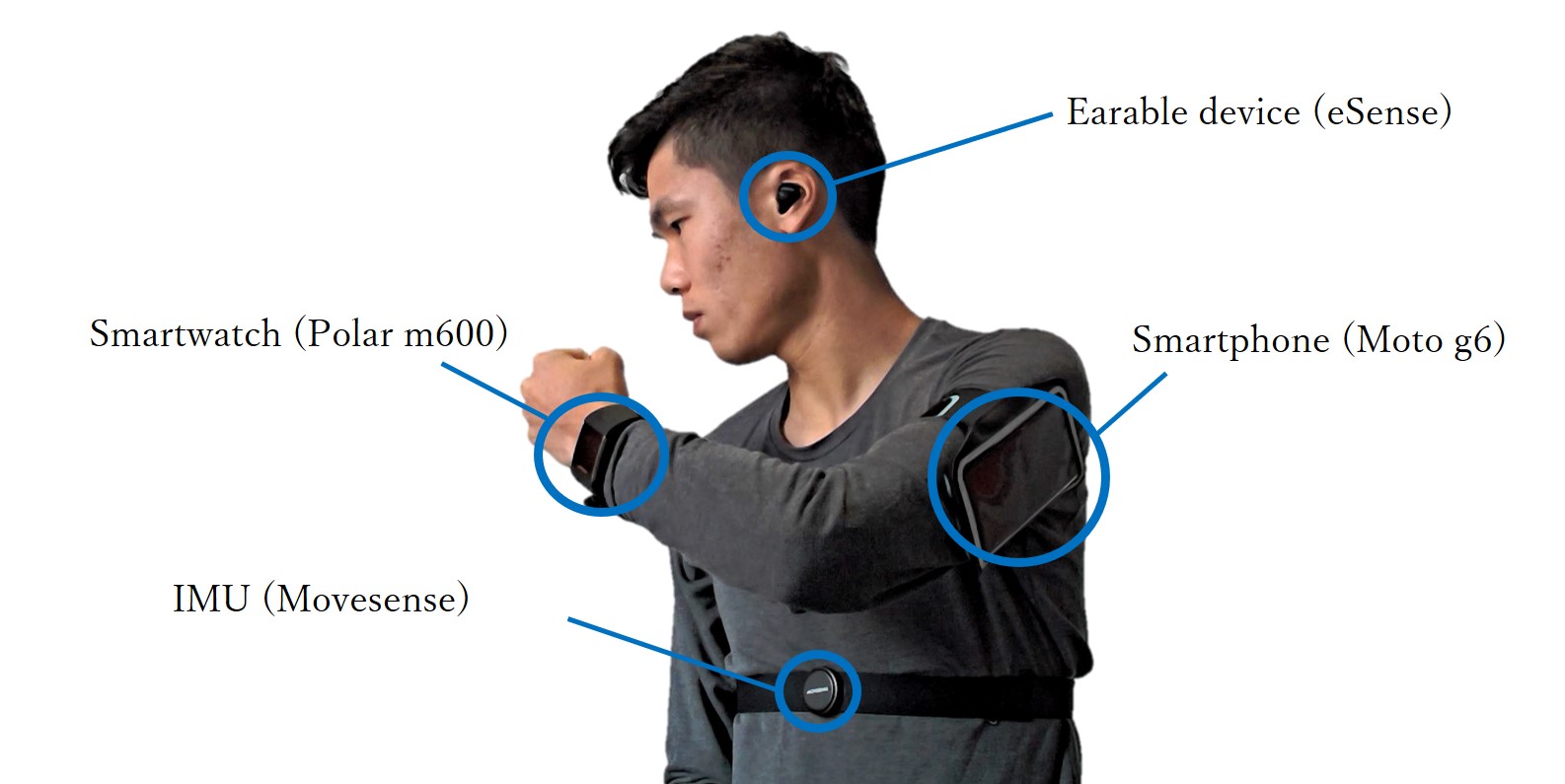}
        \caption{Sensors used for the experiment and their positionings}
        \label{sensors}
    \end{figure}
    \item \textbf{Application}\textemdash  We developed an application that can count five exercises above using the proposed method or others under development. It can also collect streamed data from a chest-mounted IMU at the same time. The application was used for this experiments.
\end{itemize}
\subsection{Definition of exercises}
The proposed method has been evaluated and validated on the following five exercises: (1) Running, (2) Walking, (3) Jumping, (4) Push-ups, (5) Sit-ups. Participants ran and walked more than 20 steps each without caring whether they start with the right or left foot. They perform jumps, push-ups, and sit-ups around ten times each. The movements of jumping, push-ups, and sit-ups were pre-defined and explained using demonstration photos because there are various kinds of movements. The author itself performed the exercises to produce the templates for all five exercises, which are necessary for real-time classification.
\subsection{Validation results of the proposed method}
\subsubsection{Confusion matrix}
Table \ref{truth_predict} shows the results of the validation of the proposed method using data collected through the experiments described in the previous section. ``Truth" corresponds to the labels that were assessed watching at the video recorded during each exercise practice. ``Predicted" corresponds to the labels that have been classified by the proposed method. ``Overlooked" corresponds to uncounted exercise even though labeled. Contrarily, ``mistook" corresponds to the number of classified labels though there was no exercise.
\begin{table}
    \centering
    \caption{Confusion matrix of Truth and Predicted labels}
    \begin{tabular}{c|c|c|c|c|c|c|c|}
        \cline{3-8}
        \multicolumn{2}{c|}{\multirow{2}{*}{}}                  & \multicolumn{5}{c|}{Predicted}                   & {\multirow{2}{*}{Overlooked}} \\ \cline{3-7}
        \multicolumn{2}{c| }{}                                   & Running & Walking & Jumping & Push-ups & Sit-ups &                             \\ \hline
        \multicolumn{1}{|c|}{\multirow{5}{*}{Truth}} & Running  & 231     & 6       & 0       & 0        & 0       & 3                           \\ \cline{2-8}
        \multicolumn{1}{|c|}{}                       & Walking  & 0       & 272     & 0       & 0        & 2       & 22                          \\ \cline{2-8}
        \multicolumn{1}{|c|}{}                       & Jumping  & 0       & 0       & 80      & 0        & 0       & 11                          \\ \cline{2-8}
        \multicolumn{1}{|c|}{}                       & Push-ups & 0       & 0       & 0       & 87       & 0       & 7                           \\ \cline{2-8}
        \multicolumn{1}{|c|}{}                       & Sit-ups  & 0       & 0       & 0       & 0        & 82      & 6                           \\ \hline
        \multicolumn{2}{|c|}{Mistook}                           & 3       & 8       & 0       & 3        & 2                                     \\ \cline{1-7}
    \end{tabular}
    \label{truth_predict}
\end{table}

\subsubsection{Segmentation results}
The precision, the recall, and the F1-score have been defined as following to evaluate the segmentation algorithm from the results of Table \ref{truth_predict}.
\begin{itemize}
  \item The precision is the fraction of predicted exercise segments that correspond to actual exercise.
  \item The recall is the fraction of actual exercise segments predicted as exercise.
  \item The F1-score is the harmonic mean of precision and recall.
\end{itemize}
Table \ref{table:segmentation} shows the results of segmentation after classification. In the segmentation step, many peaks are detected, including transition, but these no-template motions are ignored at the classification step. So, the accuracy of segmentation has been evaluated after classification. In this study, nearly 97.9\% precision and 93.9\% recall could be achieved. F1-score which shows the total accuracy of the segmentation algorithm, was 95.9\%.
\begin{table}
    \centering
    \caption{Results of segmentation algorithm performance evaluation}
    \begin{tabular}{p{2cm}p{2cm}p{1.1cm}}
        \hline\noalign{\smallskip}
        Precision & Recall & F1-score \\
        \noalign{\smallskip}\svhline\noalign{\smallskip}
        97.9\% & 93.9\% & 95.9\% \\
        \noalign{\smallskip}\hline\noalign{\smallskip}
    \end{tabular}
    \label{table:segmentation}
\end{table}
\subsubsection{Classification results}
As for the segmentation algorithm performance evaluation, the accuracy of the classification method has been evaluated the same indices: precision, recall, and F1-score. As shown in the last row of Table \ref{table:classified_exercises}, the proposed method achieved approximately 95\% despite it includes segmentation error. Besides, there was no big accuracy difference depending on the exercise, which proves the robustness of the proposed classification method based on weighted DTW scores.
\begin{table}
    \centering
    \caption{Performance results of the exercise classification method}
    \begin{tabular}{p{3cm}p{2cm}p{2cm}p{1.1cm}}
        \hline\noalign{\smallskip}
        Exercise & Precision & Recall   & F1-score \\
        \noalign{\smallskip}\svhline\noalign{\smallskip}
        Running  & 98.7\%    & 96.3\%   & 97.5\%   \\
        Walking  & 95.1\%    & 91.9\%   & 93.5\%   \\
        Jumping  & 100.0\%   & 87.9\%   & 93.6\%   \\
        Push-up  & 96.7\%    & 92.6\%   & 94.6\%   \\
        Sit-up   & 95.3\%    & 93.2\%   & 94.3\%   \\
        Micro Average & 96.9\%    & 93.0\%   & 94.9\%   \\
        \noalign{\smallskip}\hline\noalign{\smallskip}
    \end{tabular}
    \label{table:classified_exercises}
\end{table}
\par
The proposed method can count the number of times each exercise is performed, but its accuracy is depends on the segmentation and classification phase. At the counting phase, only the labeled exercises are counted, so, it is not necessary to evaluate counting performance.
\subsection{Discussion about experimental results}
The results cannot be correctly compared with previous works because the target exercises are different. Nevertheless, comprehensively, the accuracy of the proposed method is high enough to use it as actual regular exercise monitoring. Especially accuracy of approximately 95\% could be achieved, including segmentation and classification phases.
\subsubsection{Machine learning vs correlation}
Nowadays, almost all of the classification methods use machine learning because it is incredibly accurate. However, in this case, it is too difficult to collect enough data to train the model adequate to each exercise and each person. Otherwise, in the case of the proposed correlation-based approach, only one motion of the exercise is needed, and it does not require individual calibration, so, it is straightforward to collect and deploy. In case the templates are not suitable for some people to classify the exercise, they can tune it very quickly by themselves under the correlation-based method.
\subsubsection{Overlooked errors}
The proposed method classifies exercises after segmentation. In the classification phase, if the segment does not match any of the five exercises, it is ignored as a non-exercise segment. So, the overlooked segments were ignored at the classification phase because of their low similarity with the templates. In this experiment, we used our templates as correct movements. So, while we need to improve these templates of defined exercises, we can say that ``some exercises that performed by participants were not exact." To solve the problem, we need to define the exercise in more detail.
\subsubsection{Mistook errors}
On the other hand, it is assumed that mistook exercises are detected because of the following reason. The experiments included non-exercise scenes such as the transition from an exercise to another. During these transition periods, the proposed segmentation algorithm also detects peaks from the streamed acceleration signal. Most segments of the transitions are ignored at the classification phase because most of the transition movements do not correlate with the templates of each exercise. Still, few transition movements may correlate with some exercise template. From the results shown in Table \ref{truth_predict}, eight transition movements were classified into walking, three were classified into running, another three were classified into push-up, and two were classified into sit-up. The reason why some transitions were classified into walking and running is that most of the transition movements includes footsteps. Also, the reason why some transition movements were classified into push-ups and sit-ups is that the intensity of these two exercises is low. If the intensity of acceleration is low, distance scores with templates are going to low even though there is no correlation. Oppositely, there was no misclassification into jumping because it has high intensity.
\subsubsection{Running and walking}
According to the results of classification, some running steps were classified into walking. One of the reasons is because running starts from walking and ends to walking unconsciously. The first step of running is weaker than the others. Also, when stopping running, one cannot stop instantly without slowing down that proposed method classifies into walking.
\section{Conclusion and Future Work}
\subsection{Conclusion}
This paper developed a segmentation and classification method for exercise counting some exercises. Unlike most existing approaches, the proposed approach works both in indoor and outdoor environments. Conventional exercises such as running, walking, as well as some typical workouts, have been tackled. A correlation-based approach, rather than an approach using machine-leaning techniques, has been chosen developed so that the proposed method enables easy deployment to numerous different exercises. Additionally, the proposed approach  require less training data because  only one sample of motion data of each target exercises is needed.
\par
As shown in Figure \ref{table:segmentation}, the proposed segmentation algorithm achieved 98\% precision, 94\% recall, and 96\% F1-score, while the proposed classification method achieved 97\% precision, 93\% recall, and 95\% F1-score despite being affected by the errors issued from the segmentation phase. The accuracy is approximately similar or better than previous works that handled only indoor workouts and those that use a vision-based approach. These results demonstrated that our method is accurate enough to be used in actual applications.
\par
There are three types of error: overlooked, mistook, and classification errors. Overlooked is occurred when the segment has a low correlation with any of the template data of defined exercises even though the motion is detected in the segmentation phase. In order to solve that issue, it may be necessary to adjust template data and the classification method. Most classifications occurred when participants changed the exercises. Because some transitions are similar to walking, the number of mis-classifications into walking label was higher than the others. Finally, almost all of the classification errors came from running exercise. Some running steps are mis-classified into walking because running is some high intensity version of walking, and the first and last segments have low intensity.
\subsection{Challenges and future works}
There is still room for improvement of the proposed segmentation algorithm and classification method. Exercises templates were randomly selected from the data collected data by the author itself. By creating templates data from the mean of multiple data, the classification accuracy may increase. Furthermore, one of the applications of automatic exercise recognition is to be able to support people who do not like exercise even if they understand the importance of exercise for a healthier life.
\par
\begin{figure}
    \centering
    \includegraphics[width=1\linewidth]{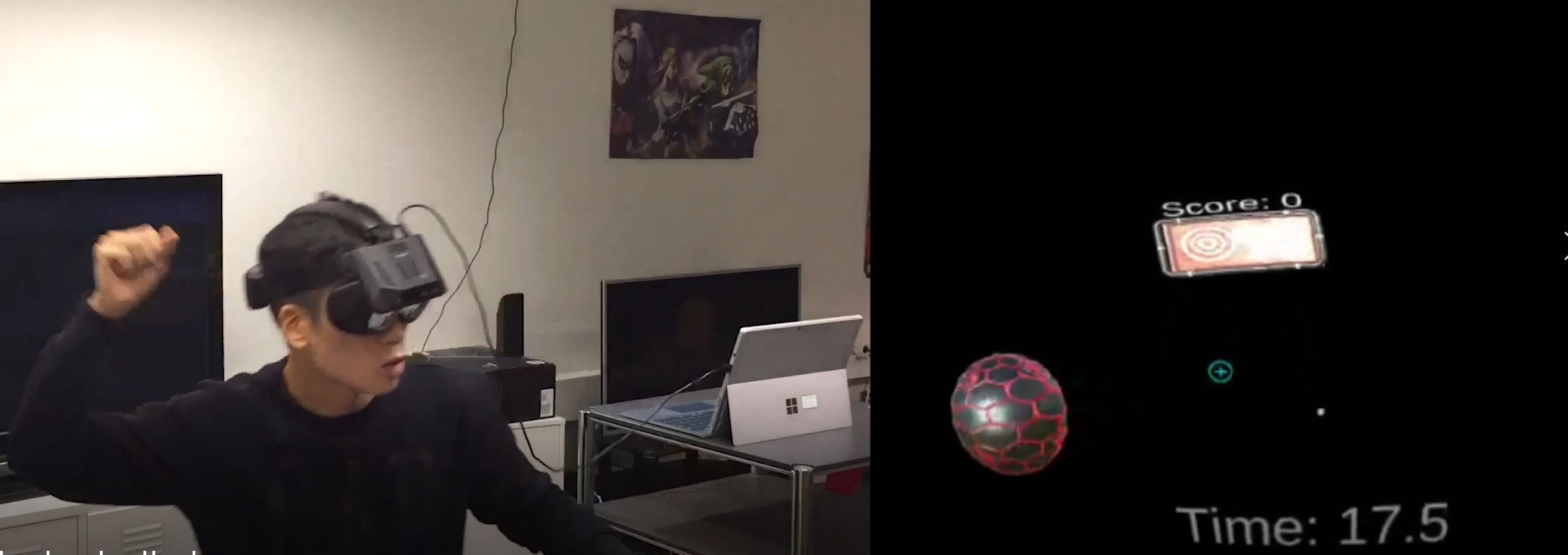}
    \caption{VR Shooting Game}
    \label{shooting}
\end{figure}
A prototype of a shooting game using a virtual reality head-mounted display has been developed to demonstrate the possibility of the technology (Figure \ref{shooting}). Using the same method proposed in this study, throwing motion can detected from the 3-D acceleration signal of an IMU attached to the wrist of the dominant hand. Throwing motion detection results in a virtual ball being thrown in the direction of the target. Also, the target throws balls towards the player, and the player has to dodge the balls. For the next step, a more exciting game combining running and some workouts will be developed, and a user study will be performed to evaluate the effect on the motivations for exercises compare with other platforms.
\bibliographystyle{unsrt}
\bibliography{reference}
\end{document}